\documentclass[prl,showpacs,preprintnumbers,twocolumn,amsmath,amssymb]{revtex4}
\usepackage{graphicx}% 
\usepackage{xcolor}
\usepackage{bm}
\usepackage{amsmath}
\usepackage{braket}

\usepackage[normalem]{ulem}
\usepackage{upgreek}

\definecolor{orange}{rgb}{1,0.5,0}

\begin{document}

\title{Spin-flip transitions induced by time-dependent electric fields in surfaces with strong spin-orbit interaction}

\author{Julen Iba\~{n}ez-Azpiroz$^{1,2}$, Asier Eiguren$^{1,2}$, Eugene Ya. Sherman$^{3,4}$, Aitor Bergara$^{1,2,5}$}
\affiliation{$^{1}$Materia Kondentsatuaren Fisika Saila, University of Basque Country UPV-EHU 48080 Bilbao, Euskal Herria, Spain}
\affiliation{$^{2}$Donostia International Physics Center (DIPC), Paseo Manuel de Lardizabal 4, 20018 Donostia/San Sebastian, Spain}
\affiliation{$^3$Department of Physical Chemistry University of Basque Country UPV-EHU 48080 Bilbao, Euskal Herria, Spain}
\affiliation{$^4$IKERBASQUE Basque Foundation for Science, 48011 Bilbao, Bizkaia, Spain}
\affiliation{$^5$Centro de F\'{i}sica de Materiales CFM - Materials Physics Center MPC, Centro Mixto CSIC-UPV/EHU,
Edificio Korta, Avenida de Tolosa 72, 20018 Donostia, Basque Country, Spain}
\date{\today}


\begin{abstract}

We present a comprehensive theoretical 
investigation of the light absorption rate at the Pb/Ge(111)$-\beta\sqrt{3}\times\sqrt{3}R30^{\circ}$ surface
with strong spin-orbit coupling. Our calculations show that electron spin-flip transitions cause as much 
as 6\% of the total light absorption, representing
one order of magnitude enhancement over Rashba-like systems.
Thus, it is demonstrated that
a substantial part of the light irradiating this nominally non-magnetic surface
is attenuated in spin flip processes.
Remarkably, the spin-flip transition 
probability is structured in well defined hot spots within the Brillouin zone  
where the electron spin experiences a sudden $90^{\circ}$ rotation.
This mechanism offers the possibility of an experimental 
approach to the spin-orbit phenomena by optical means.

\end{abstract}
\maketitle
Understanding electron spin transport and spin relaxation 
in quasi-two-dimensional (2D) systems is of a capital 
importance due to both, the fundamental reasons 
and the potential technological applications.
The spin-orbit (SO) interaction is the most prominent relativistic effect
leading to the fascinating phenomena recently observed in 2D systems,
such as the quantum spin Hall effect~\cite{spin-hall-kane,spin-hall-zhang}. 
An experimentally accessible spin degree of freedom offers 
new route for the emergent field of spintronics, 
where the main features of charge dynamics are strongly influenced 
by the spin-related effects~\cite{spin-interf-1,spin-interf-2}.
The technical possibility of spin manipulation and control
by means of an applied bias voltage is strongly supported by 
recent investigations on a variety of semiconducting 
alloy samples~\cite{gaas-so,insb-so,gate-control}.

However, a strong SO coupling cannot be achieved in conventional semiconductors 
where the spin-splitting of the conduction electrons is limited 
to a few meV at most.
As opposed, the relativistic effects completely dominate the 
electronic structure of many 
heavy-element surface materials and overlayers~\cite{highlightso}.
The reason resides on the breaking of the 
inversion symmetry and the associated 
gradient of the effective one-electron potential 
introduced at the interface.
These effects lead to 
extraordinarily large (100-500 meV) spin-splittings 
among the so called Shockley-type surface states,
as it was first observed in the 
free electron-like
Au(111) noble metal surface~\cite{lashell}. 
This system is regarded as a prototype of the 
standard Rashba model~\cite{rashba,rashba-bychkov}.
Considering the general form of the
SO interaction in the non-relativistic limit,  
\begin{equation}\label{eq:so}
\hat{H}_{\text{SO}}= -\frac{\hbar e}{4m^{2}c^{2}}
\hat{\bm{\sigma}}\cdot\left(\textbf{k}\times \bm{\nabla} V (\textbf{r})\right),
\end{equation}    
the Rashba SO coupling is recovered 
by taking the gradient of the effective one-electron 
potential $V($\textbf{r}$)$ as a constant and completely surface perpendicular.
This model produces an entirely isotropic result, with a simple 
linear spin-splitting of the two spin sub-bands with chiral spin-polarization.  
However, many heavy-element adlayer 
or even clean surface materials, such as
Bi(110) \cite{eigurenbi}, 1x1H/W(110) \cite{eigurenhw},
Au/Si(557)~\cite{ausi557}, Bi/Si(111)~\cite{bisi-exp,khomitsky}
or Tl/Si(111)~\cite{abrupt,tlsi111,minghao} 
exhibit large anisotropic SO interaction, 
inducing complex spin textures that
considerably deviate from the free electron-like picture 
of the Rashba model. 

In this Letter, we investigate electrically induced spin-flip excitations 
on the Pb/Ge(111)-$\beta\sqrt{3}\times\sqrt{3}R30\,^{\circ}$ 
surface ($\sqrt3$Pb/Ge(111)), 
considering the full spinor structure of the electron states within \textit{ab-initio} 
density functional theory. 
The goal is to understand and quantify 
the striking mechanism
leading to spin-flip transitions out of a purely electric 
perturbation in a non-magnetic material.
This system 
presents two well defined spin-spit 
surface states crossing the Fermi level, 
while the bulk substrate remains semiconducting. 
Thus, we face a problem involving a completely 
spin-polarized 2D electron gas
which is essentially decoupled from the bulk, i.e.,
an optimum scenario for 
studying surface spin-flip transitions~\cite{scontr}. 
Noteworthy, the strength of the SO interaction associated 
to the Pb atoms is two orders of magnitude 
larger ($\approx 300$ meV) than in 
semiconductor quantum wells.
Moreover, the Pb overlayer includes strong anisotropic 
gradients in the surface ionic potential, 
giving rise to a fast variation of the noncollinear 
spin-polarization in momentum space
which critically enhances 
the spin-flip transition probability.

The spin manipulation in 2D systems~\cite{tamarat_spin-flip_2008}
is accessible via 
electric-dipole spin resonance mechanism, 
which couples the spin-dependent electron velocity to externally
applied fields~\cite{spin-manipulation}.
At surfaces, the different spin-spit sub-bands are connected
by inter-band transitions that flip the electron spin~\cite{sherman_minimum}.
The detailed analysis of these processes 
provides valuable information 
about the spin-dynamics of the system, 
thus making the first-principles approach imperative. 
Simplified tight-binding Rashba-like models, although very
useful for understanding the basic physics~\cite{bisi111-2}, appear not to be realistic enough 
to account for material specific details 
which determine the electromagnetic response.

Let us consider the interaction of an electron with an 
external time-dependent electric field of frequency $\omega$.
Within the electric dipole approximation valid for small momentum transfer $\textbf{q}\to0$,
the interaction Hamiltonian is given by~\cite{spin-manipulation,sherman_minimum}
\begin{equation}\label{eq:h1v}
\hat{H}_{\text{int}}(t)= -\frac{e}{c}\hat{\textbf{v}}\cdot\textbf{A}_{\text{ext}}(t),
\end{equation}         
where $\textbf{A}_{\text{ext}}(t)=\textbf{A}_{0}\cos\omega t$ is the
external vector potential associated to the electric field 
$\textbf{E}_{\text{ext}}(t)=\textbf{E}_{0}\sin\omega t$ 
with $\textbf{E}_{0}=\textbf{A}_{0}\omega/c$,
and $\hat{\textbf{v}}$ is the electron velocity operator. 
In systems with SO coupling, 
besides the canonical contribution $\hat{\textbf{p}}/m$, 
the velocity operator includes an additional spin-dependent term
which emerges as the main responsible for the 
spin-flip transitions~\cite{sherman_minimum}.
\begin{figure}[t]
\includegraphics[width=0.35\textwidth]{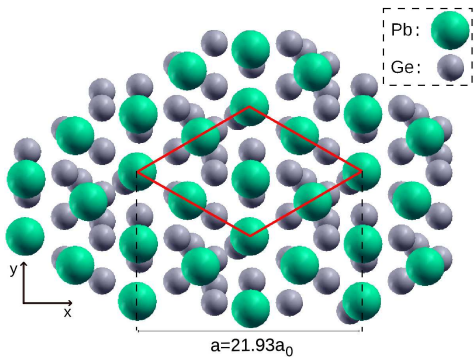}
\caption{(color online) Top view of the 
$\sqrt3$Pb/Ge(111) surface~\cite{xcrysden}. 
The small (gray) spheres symbolize the Ge substrate layers, whereas the
big (green) spheres represent the Pb surface monolayer. 
The solid (red) parallelogram indicates surface unit cell.}
\label{fig:structure}
\end{figure}

Within first order perturbation theory, 
the transition rate associated to spin-flip excitations 
between spin-spit surface states 
$S$ and $S^{\prime}$ 
due to photon absorption (term $\textbf{A}_{0}e^{-i\omega\cdot t}/2$)
can be derived from the Fermi's golden rule,
\begin{equation}\label{eq:tr-rate}
\begin{split}
\gamma_{S S^{\prime} }(\omega)&=
\frac{2\pi}{\hbar}
\int_{SBZ} \Big(f(\epsilon_{S{\bf k}})-f(\epsilon_{S^{\prime}{\bf k}})\Big)
 \\
&\left| M_{S  S^{\prime}}(\textbf{k})\right|^{2}
\delta(\epsilon_{S{\bf k}}-\epsilon_{S^{\prime}\textbf{k}}-\hbar\omega)\frac{d^{2}{k}}{(2\pi)^{2}}.
\end{split}
\end{equation}
The integral is taken over the surface Brillouin zone (SBZ), 
$f(\epsilon_{i,{\bf k}})$ and $\epsilon_{i,\textbf{k}}$ 
represent the Fermi-Dirac distribution and a surface state eigenvalue, 
respectively, and $M_{S  S^{\prime}}(\textbf{k})$ is 
the interband matrix element
\begin{equation}\label{eq:sp-prob}
\begin{split}
M_{S  S^{\prime}}(\textbf{k})=-\frac{e\textbf{A}_{0}}{2c} \cdot\bra{\Psi_{S\textbf{k}}} \hat{\textbf{v}}\ket{\Psi_{S^{\prime}{\bf k}}}.
\end{split}
\end{equation}
Above, $\Psi_{i\textbf{k}}(\textbf{r})$ denotes the 
single-particle Bloch spinor wave function 
associated to a surface state.

\begin{figure}[t]
\includegraphics[width=0.45\textwidth]{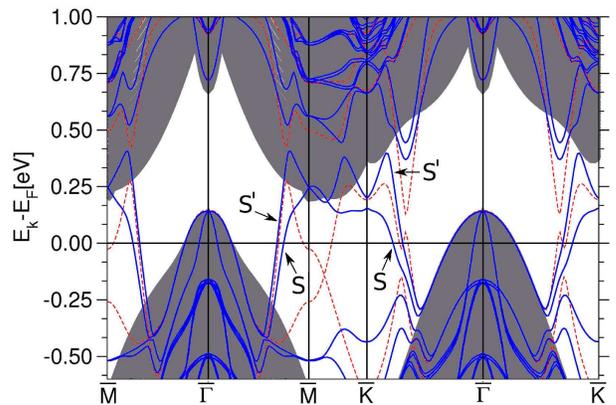}
\caption{(color online) Electron band structure of the 
$\sqrt3$Pb/Ge(111) surface. 
The scalar and fully relativistic bands are represented by 
dashed (red) and solid thick (blue) lines, respectively. 
The continuous background is the bulk band projection. 
The fully relativistic metallic 
surface states are labeled as $S$ and $S^{\prime}$.}
\label{fig:band-str}
\end{figure}

Due to the inherent phase indeterminacy of the Bloch states in \textbf{k}-space, 
the matrix elements of the velocity operator require a special treatment~\cite{blount}.
Following the approach presented in Ref. \cite{k-gradients}, we expressed 
$M_{S  S^{\prime}}(\textbf{k})$
in Eq. \ref{eq:sp-prob}
in terms of the so called maximally localized Wannier functions~\cite{wannier-entangled}.
In this way, 
the matrix elements entering 
Eq. \ref{eq:tr-rate} are maximally smooth
and thus suitable for any interpolation procedure within the Brillouin zone.
The Wannier functions were generated considering 
the entire structure of \textit{ab-initio} spinor wave functions obtained
within the noncollinear DFT formalism~\cite{wannier90,espresso,Dalcorso}.
The exchange-correlation energy was approximated by the standard LDA-PZ parametrization~\cite{lda}
and the 2$\times$2 norm-conserving fully relativistic pseudopotential approach~\cite{bylander,Dalcorso}.
The ground state self-consistent cycle was performed 
considering the usual Monckhorst-Pack mesh 
corresponding to a $27\times27$ grid. 
We employed a very fine $200\times200$ mesh
considering the standard Wannier interpolation procedure 
for all the ingredients entering Eq. \ref{eq:tr-rate}~\cite{k-gradients,wannier-entangled}
in order to reliably account for the details close to the Fermi level~\cite{k-gradients}.

\begin{figure*}[t]
\centering
\includegraphics[width=0.8\linewidth]{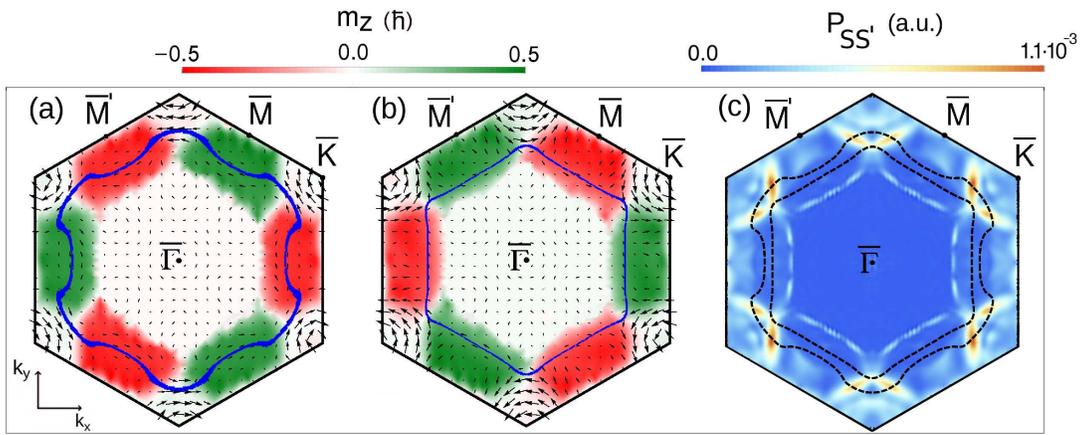}
\caption{(color online) (a) and (b) Momentum dependent spin-polarization structure 
associated to 
the $S$ and $S^{\prime}$ surface states, respectively. 
Arrows represent the in-plane spin-polarization component, 
whereas the background indicates the surface perpendicular component 
of the magnetization, $m_{z}(\textbf{k})$. The Fermi surface of each state 
is indicated by solid (blue) lines.
(c) Spin-flip transition probability 
associated to the $S$ and $S^{\prime}$ surface states
for R-circularly polarized light. 
}
\label{fig:spin-flip}
\end{figure*}

The $\sqrt3$Pb/Ge(111) surface 
(Fig. \ref{fig:structure}) was simulated by a repeated slab technique containing 
14 Ge layers. The Pb monolayer was included only in one side of the slab, 
while the other (bare) Ge(111) surface was covered by a hydrogen 
adlayer in order to saturate the dangling bonds.
We also analyzed the Au(111) metal surface 
considering a 22 Au layer slab.
In both systems, a full geometry optimization was performed 
until all atomic forces exerted on individual atoms 
were negligibly small ($<10^{-4}$ Ry a.u.$^{-1}$).

Fig. \ref{fig:band-str} shows the calculated band structure of the
Pb/Ge(111)$\sqrt{3}$ surface.
While the scalar relativistic calculation shows a single
spin-degenerate surface band crossing the Fermi level, fully relativistic calculations
present two spin-polarized surface bands labeled as $S$ and $S^{\prime}$. 
The SO interaction has a huge impact on 
the electron structure close to the Fermi level, such that
this term cannot be treated perturbatively. Its 
contribution is even more important than some non-relativistic DFT terms
such as the exchange-correlation energy.
In this context, the SO interaction completely 
determines the metallic character of the $S$ and $S^{\prime}$ surface states.
These exist as surface states only outside the area
close to the $\overline{\Gamma}$ point, where 
$S$ and $S'$ become resonances entering the bulk projection
(continuum in Fig. \ref{fig:band-str}).
Outside this region, the splitting is in overall of the order of 100 meV, reaching a
maximum of 300 meV near the high symmetry point $\overline{M}$.
The calculated Fermi wave vectors along the high 
symmetry direction $\overline{\Gamma} \overline{\text{M}}$, $k^{S}_{F}\simeq 0.41$ \AA$^{-1}$ and 
$k^{S^{\prime}}_{F}\simeq0.37$ \AA$^{-1}$, are in very good agreement 
with recent ARPES experiments reporting
$k^{ S}_{F}= 0.40$ \AA$^{-1}$ and $k^{ S^{\prime}}_{F}=0.36$ \AA$^{-1}$~\cite{scontr}. 

A characteristic feature emerging from the SO interaction at surfaces is the
momentum-dependent spin polarization:
\begin{eqnarray}
\label{eq:spin-pol} 
\boldsymbol{m}_{n}({\bf k})=\int \Psi^{*}_{n{\bf k}}({\bf r})  
\hat{\boldsymbol{\sigma}} \Psi_{n{\bf k}}({\bf r}){d}^{3}{r}, 
\end{eqnarray}
where $n$ is the band index.
In Figs. \ref{fig:spin-flip}a and \ref{fig:spin-flip}b we present 
the calculated spin-polarization for the $S$ and $S^{\prime}$ surface states in the 
entire Brillouin zone. 
These figures demonstrate that the $S$ and $S^{\prime}$  
states are spin-polarized in almost the opposite direction, 
in agreement with recent spin-resolved ARPES measurements~\cite{scontr}. 
The negligible spin-polarization around $\overline{\Gamma}$ 
is consistent with the overlap of the surface bands with the bulk projection
(see Fig. \ref{fig:band-str}).
In this area, the 
electron states become resonances with 
a large penetration, 
thus any surface effect such as the 
enhancement of the SO interaction is almost completely absent.

The anisotropic character of the 
SO interaction is evidenced by the highly noncollinear structure of 
the calculated spin-polarization for $S$ and $S^{\prime}$. 
On one hand, we observe that the spin 
of each surface state
is mainly polarized along the surface perpendicular direction, a phenomenon
that extends to a significant area around the high symmetry points 
$\overline{\text{M}}$ and $\overline{\text{M}^{\prime}}$. 
Such an important contribution of the out-of-plane magnetization 
is a consequence of the strong in-plane gradients of the ionic potential,
as reported in the Tl/Si(111) surface~\cite{minghao,abrupt,tlsi111}.
On the other hand, our calculations further identify 
an important area of almost pure in-plane 
circular spin-polarization for each state
around high symmetry point $\overline{\text{K}}$.

Fig. \ref{fig:spin-flip}c shows the results for the calculated 
spin-flip transition probability 
$P_{S  S^{\prime}}(\textbf{k})\equiv|M_{S  S^{\prime}}(\textbf{k})|^{2}/|\textbf{A}_{0}|^{2}$
for an R-circularly polarized external field,
$\textbf{A}_{0}=A_0(\hat{\textbf{x}}+i\hat{\textbf{y}})/\sqrt{2}$. 
Results for other polarizations, although being slightly different, 
do not substantially modify our general conclusions.
We deduce from Fig. \ref{fig:spin-flip}c that the spin-flip transition probability is negligible
around the $\overline{\Gamma}$ point. 
This is consistent with Eq. \ref{eq:sp-prob} since the surface states 
are spin-degenerate here,~\cite{k-gradients} therefore SO driven effects such 
as the spin-flip excitations are weak.
On the other hand, the most important message of Fig. \ref{fig:spin-flip}c is 
the extreme localization of the spin-flip transition probability 
in hot spots close to the high symmetry point $\overline{\text{K}}$.
A careful comparison of Figs. \ref{fig:spin-flip}a and \ref{fig:spin-flip}b
with Fig. \ref{fig:spin-flip}c reveals that the hot spots are localized
in the boundaries separating the surface-perpendicular and surface-parallel 
spin-polarized regions in the Brillouin zone.
This feature is fully consistent with Eq. \ref{eq:sp-prob} since the
velocity operator introduces a \textbf{k}-space derivative~\cite{blount,k-gradients}
which effectively measures the variation of the entire wavefunction
and the spin-polarization through Eq. \ref{eq:spin-pol}.
Noteworthy, this singular effect is 
completely absent in Rashba-like systems exhibiting a smooth behavior of the 
spin-polarization~\cite{rashba,rashba-bychkov,lashell}.

The light absorption rate 
associated to surface spin-flip excitations 
is given by the following expression,
\begin{eqnarray}\label{eq:absorption}
\Omega_{S  S^{\prime}  }(\omega)=\frac{\hbar\omega\cdot\gamma_{S  S^{\prime}  }(\omega)}{W}.
\end{eqnarray}
Above, $\gamma_{S  S^{\prime}  }(\omega)$ is the spin-flip 
transition rate (Eq. \ref{eq:tr-rate}), 
$\hbar\omega$ the energy of the external field and $W=c|E_{0}|^{2}/8\pi$,
the incident optical power per unit area.

\begin{figure}[t]
\includegraphics[width=0.5\textwidth]{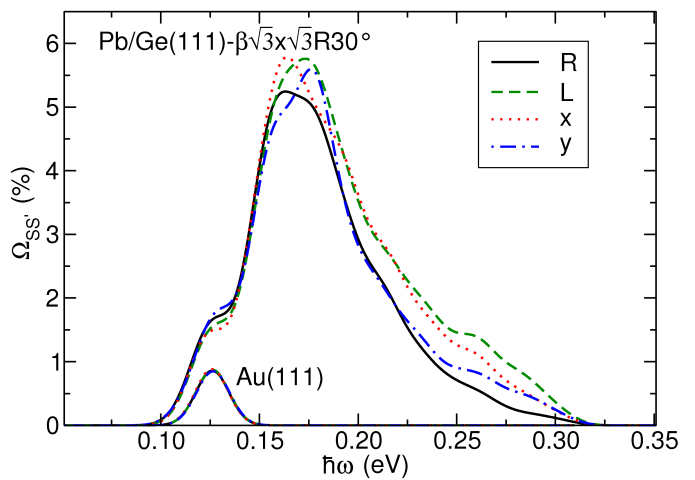}
\caption{(color online) Calculated spin-flip absorption rate in 
$\sqrt3$Pb/Ge(111)
and Au(111).
Solid (black), dashed (green), dotted (red) and dashed-doted (blue) lines 
represent the results corresponding to the
R and L circularly polarized and $x$ and $y$ linearly polarized light, respectively.
Note that the absorption rate for 
$\sqrt3$Pb/Ge(111)
depends on the light polarization, 
while in Au(111) it is practically polarization-independent. 
}
\label{fig:ab-rate}
\end{figure}

Fig. \ref{fig:ab-rate} illustrates the calculated absorption rate
associated to the spin-spit surface states of the $\sqrt3$Pb/Ge(111) and 
the Rashba-like Au(111) surfaces.
For $\sqrt3$Pb/Ge(111), the absorption spectrum is bounded
in the 0.1-0.3 eV energy range, corresponding approximately to the spin-splitting
at the Fermi level. 
At this surface, the spectrum presents a prominent peak 
close to 0.17 eV, where the spin-flip absorption rate reaches a remarkable 
maximum value of $6\%$. 

This result demonstrates that a significant 
part of the incoming radiation is dissipated exclusively in the
spin-flip phenomena.
The SO coupling making the orbital and spin degrees
of freedom interrelated causes a question on the angular momentum conservation.
As the absorption of circularly polarized light induces a net transfer of angular momentum, 
the resulting magnetization of the surface states can produce a corresponding angula momentum reservoir in a spin-orbit coupled system.
Due to the anisotropy of the spin-polarization (see Fig. 3), 
the absorption spectrum exhibits a substantial variation as a function
of the light polarization.
As an example, close to $0.25$ eV the spin-flip absorption rate for L polarized light
is approximately three times stronger than for R polarized light.
In contrast, we find that the absorption spectrum of 
Au(111) is practically polarization
independent. The reason is that this 
system exhibits an almost perfect isotropy,
as assumed in the Rashba model.
The magnitude of the spin-flip contribution for Au(111) is relatively weak
($0.8$\% maximum). 
Thus, the spin-flip absorption in $\sqrt3$Pb/Ge(111) is one order of magnitude stronger
in comparison to Au(111).

It is noteworthy that the bare spin-flip contribution to 
the absorption rate in $\sqrt3$Pb/Ge(111) is around three times stronger 
than the total absorption of a graphene layer ($2.3$\%), where the electron spin does
not play any significant role~\cite{graphene-conduct,graphite-conduct}.
Thus, the \textit{a priori} 
weaker relativistic SO interaction in $\sqrt3$Pb/Ge(111) exceeds 
the effect of the usually predominant non-relativistic terms such as the electric dipole mechanism.
The reason why the spin-flip contribution in this system
is so important is that the hot spot matrix elements (Fig. \ref{fig:spin-flip}c)
lie just inside the Brillouin zone area where the $S$ state is occupied and $S^{\prime}$ state remains empty.
In this way, the Fermi occupation factors allow electron transitions
precisely where the matrix elements are maximal.
From the discussion above we can conclude that a large anomalous feature associated 
to the enhanced spin-flip excitation mechanism should be accessible 
by infrared optical spectroscopy in the $0.1-0.3$ eV energy range.

In summary, we investigate the role of the spin-orbit interaction
on the light absorption rate of the $\sqrt3$Pb/Ge(111) surface.
Our calculations incorporate the full spinor wave function structure from first-principles, making use of a 
precise integration procedure through a Wannier interpolation scheme for the spin-flip matrix elements.
We find that a substantial part of the low-energy absorption
spectrum is dominated exclusively by the spin-flip excitations 
associated to the spin-polarized surface states. 
Noteworthy, these transitions capture as much 
as $6$\% of the total incident power,
representing an enhancement of one order of magnitude in comparison to
the Rashba-like prototype Au(111) surface.
The origin of such a huge absorption rate is closely related
to the strong anisotropy exhibited by the 
spin-polarization structure connected
to the surface states.
As demonstrated in this Letter, a clear fingerprint of the spin-flip 
absorption mechanism should be accessible in the optical range,
posing a challenge for further experimental work.

We acknowledge fruitful discussions with B. Rousseau, I. Errea, A. Kuzmenko and R. Winkler. 
This work was supported by the UPV/EHU (Grant No. IT-366-07 and program UFI 11/55), 
the Spanish Ministry of Science and Innovation (Grants No. FIS2010-19609-C02-00 and No. FIS2009-12773-C02-01), 
and the Basque Government (Grant No. IT472-10). 
The authors also acknowledge the Donostia International Physics Center (DIPC) for providing the computer facilities.

%\bibliographystyle{unsrt}
%\bibliographystyle{apsrev}
%%%%%%%\bibliographystyle{prsty}
%\bibliography{biblio}

\end{document}